# Smoking Supernovae

H L Gomez [1], S A Eales [1], L Dunne [2]

1. School of Physics and Astronomy, Cardiff University, Queens Buildings, The Parade, Cardiff, Wales, CF24 3AA, UK, email: haley.Gomez@astro.cf.ac.uk
2. School of Physics and Astronomy, University of Nottingham, University Park, Nottingham, NG7 2RD, UK

## Abstract

The question "Are supernovae important sources of dust?" is a contentious one. Observations with the Infrared Astronomical Satellite (IRAS) and the Infrared Space Observatory (ISO) only detected very small amounts of hot dust in supernova remnants. Here, we review observations of two young Galactic remnants with the Submillimetre Common User Bolometer Array (SCUBA), which imply that large quantities of dust are produced by supernovae. The association of dust with the Cassiopeia A remnant is in question due to the contamination of foreground material. In this article, we compare the emission from cold dust with CO emission towards Kepler's supernova remnant. We detect very little CO at the location of the submillimetre peaks. A comparison of masses from the CO and the dust clouds are made, and we estimate the 3 sigma upper limit on the gas-to-dust ratios to range from 25 - 65 suggesting that we cannot yet rule out freshly-formed or swept up circumstellar dust in Kepler's supernova remnant.

## Introduction

Interstellar dust plays an important role in astronomy, yet we know relatively little about the origin and evolution of the dust cycle in galaxies. Although dust grains only constitute around 1 % of the mass of the interstellar medium, they affect our view of the Universe by blocking out optical light and changing the visible appearance of astrophysical objects. Dust grains scatter, absorb and re-emit light to longer wavelengths so effectively that observing techniques at wavelengths other than optical are needed to obtain a complete picture of the Universe. Recently, astronomers have realised that dust is far more than the 'smoke' between the stars, particularly since the advent of infra-red and sub-millimetre telescopes which directly detect the recycled emission from dust grains. Surveys have shown that dust plays an important part in the cooling processes of the gas and its interaction with the gas dynamically, as well providing greater understanding of stellar chemistry; indeed, it is believed to be the main catalyst for the formation of molecular hydrogen in space (Hirashita & Ferrara 2002), an effective coolant in star forming regions, an important tracer of metals in the Universe (Dunne et al. 2003a) and a possible tracer for high density gas. Perhaps the most convincing argument for the importance of understanding the origin and evolution of dust is seen in the recent studies of the infrared and submillimetre background. These observations clearly show that the amount of energy in the infrared/submillimetre background is almost as much as that in the optical background (Fig 1). This has one serious implication – almost half of all the optical light emitted since the Big Bang has been absorbed and re-radiated by dust.

Although we recognise the importance of interstellar dust, even the source of dust in the Galaxy is unknown. There are many observations which provide evidence for dust grain formation in stellar outflows: infra-red emission around red giants, planetary nebulae, Wolf-Rayet stars and carbon stars. Indeed, stellar winds (SWs) are thought to be the most important contributors to stardust into the ISM (Whittet 2003; Jones et al. 1996; Draine 2003). The question is, *how much*? Dust production in stars is hard to quantify observationally, it depends on the mass of heavy elements in the stellar atmospheres and the mass loss rates during the final stages of the star's evolution. The required cycle to produce dust in stars begins with enrichment of the ISM from the first population of rapidly evolving supernovae (SNe), the incorporation of these elements into star formation and the evolution of the stars to the right atmospheric conditions before significant dust production can occur. The timescales for dust injection from stars is of the order of a few billion years. Observations of mass loss rates from intermediate mass stars suggest that they contribute 86 – 97 % of the total dust mass injected from astrophysical sources (Whittet 2003). However, there is a major problem with this statement: there is not enough dust in SWs to explain the mass of dust we see in our own Galaxy (known as the dust budget crisis). This problem is further compounded with the recent discovery of a population of extremely dusty objects at high redshifts seen in blank field submillimetre surveys and observations of distant quasars (Bertoldi et al. 2003; Eales et al. 2003) which imply that dusty galaxies are present in the Universe at $z > 4$ (Smail et al. 1997; Isaak et al. 2002). The Universe was less than 1/10th of its present day age at this time and it is difficult for the dust to have originated from the stellar winds of intermediate mass stars in such short timescales. An alternative source of dust could be supernovae, as they provide large abundances of heavy elements and can create the required density/temperature/pressure conditions for dust to condense (Clayton et al. 2001; Todini & Ferrara 2001; Nozawa et al. 2003). Type-II supernovae are the explosions of massive stars, which evolve rapidly and reach the supernova phase after only 10 - 100 Myr. Thus Type-II supernovae could potentially provide a rapid source of dust. If there is little or no dust formation from rapidly-evolving supernovae then it is difficult to understand where the high redshift dust originated. This problem is highlighted in Fig 2 where the evolution of dust mass for a galaxy ($10^{10}$ $M_{solar}$) with time is shown using a theoretical chemical evolution model to show how the dust builds up with time (Morgan & Edmunds 2003). The two solid lines represent dust mass from supernovae (SNe) and stellar winds (SW). If SNe are not important contributors to the interstellar dust budget, it will take this galaxy at least 5 billion years ($z < 2$) to build up a dust mass of $> 10^7$ $M_{solar}$ from stellar winds only. The mass of dust seen in high-$z$ galaxies and quasar systems at $z > 4$ is around $10^8$ $M_{solar}$. These problems suggest that dust formation in supernovae (or, more importantly, a rapid source of dust) is required to explain both the presence of high redshift dust and the dust mass in our own Galaxy.

Theorists have long championed dust formation in supernovae (e.g. Clayton et al. 2001, Todini & Ferrara 2001, Nozawa et al 2003) yet observations with infrared cameras such as IRAS showed very little amounts of dust observationally. In one galactic supernova remnant (SNR), Cassiopeia A, astronomers observed a tiny $10^{-7}$ $M_{solar}$ of dust with IRAS. Later observations with ISO hinted at the presence of much more dust, finding 0.15 $M_{solar}$ (Tuffs et al. 1999), but the low resolution of ISO and instrumental difficulties at the longer wavelengths, meant that this result was largely ignored. The total dust yield from SNe estimated using the observations, is no more than $10^{-5}$ $M_{solar}$ of dust per year, whereas the models imply that they could inject anywhere between $(0.4 – 40) \times 10^{-3}$ $M_{solar}$ of dust per year. The huge discrepancy between the theoretical models and observations of supernovae could be explained if there existed a population of cold dust in the supernova ejecta not visible with infrared telescopes, which are sensitive to emission from hot dust only. Such a population of cold dust grains would emit at longer, submillimetre (submm) wavelengths. The Submillimetre Common User Bolometer Array (SCUBA) camera on the James Clerk Maxwell Telescope (JCMT) in Hawaii was used to observe two young supernova remnants, Cas A and Kepler, at 850 and 450 μm. (To determine if the dust is freshly formed by the supernova explosion or blast wave, we require observations of 'young' remnants that are still dominated by the ejecta dynamics i.e. have not swept up much gas. This limits us to very small numbers of possible sources, further compounded by poor sensitivity with SCUBA and that most of these objects lie towards the centre of the Galaxy and are confused by foreground material.) These observations discovered large amounts of submm emission, suggesting three orders of magnitude more dust existed in the remnants than seen with IRAS and ISO. Recent observations have suggested that there may be alternative explanations for the large dust masses detected with SCUBA.

In this paper, we review the original submm observations and data reduction of these two remnants, including a detailed description of how the dust mass was estimated. We compare the submm observations with our new carbon monoxide (CO) images towards the remnants to determine if there is contamination from foreground molecular clouds. We also review whether or not the SCUBA dust could be from an 'exotic' form of dust grains, in the shape of iron needles. Finally we discuss the consequences of the CO and submm observations.

## Submillimetre Observations of Supernovae- Cassiopeia A

Cas A is the brightest radio source in the sky and is believed to be the remnant of a massive star which exploded around 300 years ago. It lies at a distance of approximately 3.5 kpc with diameter of approximately 8 arcmins. The SCUBA observations were made in 1995 in the scan-mapping mode and were available in the JCMT archive. After data reduction, low-level regions of diffuse emission remained on the image. Scan-mapping typically leaves such artifacts and they vary depending on the methods chosen for removing the baselines. Therefore a surface was fitted to the image, which left the background flat. This was checked by taking extra photometry data (not prone to the same systematics as scan mapping) in Dec 2002 at both 850 and 450 μm, which provided an independent check of the absolute flux levels at several positions on the remnant.

The SCUBA images of Cassiopeia A are shown in Fig 3 at (a) 850 μm with 450 μm contours overlaid (3σ + 1σ) and (b) 450 μm (Dunne et al 2003b). Around two-thirds of the emission at 850 μm is contaminated with synchrotron emission described by a power law slope $\nu^{-\alpha}$. Once this component is subtracted, we can see the emission from cold dust only; Fig 3(a) shows the synchrotron-subtracted 850 μm image. The forward and reverse shock fronts seen in the X-ray from the supernova blast wave are also overlaid (Gotthelf et al 2001). Once the synchrotron is subtracted the 850 and 450 μm emission follow a similar distribution with the cold dust now located mainly in the south and eastern parts of the remnant. The submm peaks appear to fall between the forward and reverse shocks seen in Fig 3(a). The final integrated submm fluxes for Cas A's SNR, minus the synchrotron component (i.e. from cold dust) are $S_{850} \sim 15.8 \pm 5.6$ Jy and $S_{450} \sim 47.5 \pm 16.1$ Jy.

## Kepler

The explosion in 1604 left behind a shell-like remnant of approximately 3 arcmin in diameter which lies at a distance of ~ 5 kpc. The progenitor and supernova type is controversial (Schaefer 1996, Blair 2004). There is dynamical evidence to suggest that the explosion was a Type II (massive star explosion) along with an overabundance of nitrogen thought to be made from the CNO cycle of massive stars (e.g. Borkowski et al. 1992), but model-fitting to X-ray spectra suggest the ejecta has chemical composition similar to that expected from a Type Ia explosion (the nuclear explosion from a white dwarf binary system, e.g. Kinugasa & Tsunemi (1999)).

Six 'jiggle-map' observations were centred around the SNR since the remnant is larger than SCUBA's field-of-view, with chop throw 180 arcsec. Figure 4 shows the SCUBA signal-to-noise images of Kepler at (a) 850 μm and (b) 450 μm (Morgan et al. 2003); the shell-like structure is clearly visible at 850 μm. The synchrotron component is far less in Kepler as it is not as radio bright as Cassiopeia A. The final integrated submm fluxes for Kepler's SNR, minus the synchrotron component are $S_{850}$ ~ 1.0 ± 0.16 Jy and $S_{450}$ ~ 3.0 ± 0.7 Jy.

## Estimating the Dust Mass

The dust mass can be measured directly from the flux at submm wavelengths using (Hildebrand, 1983)

$$M_d = \frac{S_\nu D^2}{\kappa_\nu B(\nu, T_d)}$$

where $S_\nu$ is the flux density measured at frequency $\nu$, D is the distance and $\kappa_\nu$ is the dust mass absorption coefficient. $B(\nu, T)$ is the Planck function and $T_d$ is the dust temperature. We fitted a two-temperature greybody to the infrared – submm spectral energy distribution (SED) of the two remnants, allowing the dust emissivity parameter, β and the warm and cold temperatures to vary. The best-fit SED for Kepler and Cassiopeia A are shown in Figs 5 and 6 respectively, with the best-fit parameters for each SNR listed in the captions. We used a bootstrap technique to derive errors on these values, creating 3000 sets of artificial fluxes from the original fluxes and their associated error bars. Our two-temperature model was then applied to each artificial set and errors derived from the distribution of $T_{warm}$, $T_{cold}$ and β produced by these fits (inset).

The largest uncertainty in the dust mass comes from the uncertainty in $\kappa_\nu$. We have followed Dunne et al. (2003a) in trying three different values of $\kappa_\nu$ from the literature: (1) $\kappa_{850\mu m}$ ~ 0.85 m² kg⁻¹, the average value from the range observed in laboratory studies of clumpy aggregates; (2) $\kappa_{850\mu m}$ ~ 0.48 m² kg⁻¹, the average observed in circumstellar environments and (3) $\kappa_{850\mu m}$ ~ 0.01 m² kg⁻¹, the average observed for the

*diffuse* ISM where dust is likely to have encountered extensive processing. In Cas A, the higher $\kappa_{850\mu m}$ values were required to give a reasonable dust mass of **2.6 ± 0.7 M$_{solar}$**. If we used the κ values relevant for 'normal' interstellar dust, the dust mass is uncomfortably large, greater than **15.0 ± 4 M$_{solar}$**. Using the laboratory κ values for Kepler gives a lower limit of **0.3 ± 0.1 M$_{solar}$**, whereas using the 'normal' dust κ values, gives **2.7 ± 0.6 M$_{solar}$**.

## Alternative Explanations

Given the importance of determining the correct mass of dust produced by supernovae or their massive star progenitors, two competing theories have been put forward claiming that the dust mass in these remnants is in fact much lower: (i) 'exotic' needle-like metallic grains are responsible for the dust emission and (ii) the emission in Cas A is contaminated by foreground material and is not associated with the remnant. In this section, we discuss the evidence for and against both theories and their possible implications.

### 'Exotic' Dust Grains

Conducting iron needles were proposed as an alternative explanation for the emission in the SCUBA image of Cas A (Dwek 2004). Such needles, if they exist, would be efficient emitters at submillimetre wavelengths and would be collisionally heated by the hot X-ray gases in the supernova blast wave to temperatures of around 10 K. The high emissivity of the needles gives rise to large absorption coefficients, which serve to decrease the dust mass determined from the emission by several orders of magnitude (e.g Edmunds & Wickramasinghe 1975). The dust mass absorption coefficient for iron needles is given by

$$\kappa = \frac{4\pi}{3c\rho_d\rho_r}$$

where $\rho_d$ is the density of iron and $\rho_r$ is the conductivity. The variation of the absorption coefficient of an iron needle with axial ratio (length/radius, $l/a$) for radius = 0.1 μm at 450 μm is show in Fig 7(a). The variation of the absorption coefficient of iron needles with wavelength for different axial ratios is plotted in Fig 7(b). The needles are modelled as antenna with resistivity $10^{-5}$ Ω cm. The absorption coefficient at 850 μm for iron needles with $l/a \sim 10,000$ is $\sim 10^5$ m$^2$ kg$^{-1}$. For comparison, the absorption coefficient at 850 μm for 'normal' interstellar dust is $\sim 0.07$ m$^2$ kg$^{-1}$. Given that $M_d \propto 1/\kappa$, Dwek (2004) estimated that the mass of iron needles required to explain the submm emission from Cas A would only be $\sim 10^{-5}$ M$_{solar}$. Using Dwek's formulisation of the heating and cooling of the needles in the SN blast wave, we investigated whether or not the emission from Kepler's remnant could be explained by these 'exotic' dust grains (Gomez et al 2005). We found that the mass of iron needles required to explain the submm emission from Kepler would be $< 10^{-3}$ M$_{solar}$. In this case, we no longer have a significant source of dust in the early Universe, although if the dust in the high-z galaxies are also composed of iron needles, the galactic dust mass would also decrease.

Using this model, we found that the parameters required to fit the SED and observed properties of Kepler is inconsistent with that suggested for Cas A. An additional, more serious problem with the iron needle model is that it is based on the Rayleigh criterion (Li 2003), which needs to be satisfied to produce the absorption efficiencies seen in Figure 7. Using the range of axial ratios ($l/a < 700$) and conductivities ($\rho_r \sim (4 - 60) \times 10^{-17}$ s) required to fit Kepler's SED, the Rayleigh criterion is only satisfied for iron needles with grain radii of 0.8 – 5.7 Å. This is equivalent to approximately a few layers of iron atoms at most. It is extremely difficult to explain how such small grains with length 1000 times greater than their radius would form and indeed survive in the hot X-ray plasma.

**Foreground Interstellar Clouds**

Dunne et al. (2003b) assumed that the SCUBA dust was associated with Cas A for a number of reasons. The strongest evidence for this assumption is that the 850 μm emission is completely bounded by the forward and reverse shocks of the remnant (as determined by the X-ray and radio observations). Second, they compared the submm emission with the available CO maps of the remnant in the literature (e.g. Wilson et al 1993; Liszt & Lucas 1999) and found very little evidence for a correlation between the SCUBA peaks and the CO maps. These CO observations indicated highly diffuse emission over the entire remnant with a stronger concentration in the south. The submm emission is not diffusely distributed outside the remnant in the same manner as the CO and is clumpy on small scales. Finally, they estimated the dust mass in the CO peaks using a gas-to-dust ratio of 150:1 and found dust masses much lower than the submm emission predicted. We subsequently obtained our own CO maps of Cas A with the A3 receiver on the James Clerk Maxwell Telescope (JCMT) in 2004, as part of the JCMT Service Programme (see Fig 8). The CO emission has been integrated over the velocity interval $-50 < v < -35$ km/s, which includes all the gas from the Perseus spiral arm. Overlaid are SCUBA contours at (a) 850 μm with synchrotron emission subtracted and (b) 450 μm. Some of the peaks in the submm continuum are at the positions of peaks in the molecular gas, which suggests that some of the dust may be foreground material or dust that has been swept up by the blast wave. Notice also the correlation between CO and SCUBA dust clumps well outside the remnant (clumps A - D). These clumps are distributed in a ring like structure centred on the remnant with radius of 290 arcsec (4.8 pc) and individual sizes of 0.5 - 1 pc across. The origin of these clumps is not yet understood but could have been formed in the progenitor's stellar wind or high velocity ejecta clumps from the supernova.

Wilson & Bartla (2005) used CO observations towards Cas A to show positional agreement for three of the submm peaks with CO clouds and estimated that *half* of the dust in the remnant could be associated with the intervening interstellar clouds. This leaves around 1 $M_{solar}$ of dust in Cas A (10 000 times more than detected with previous far-infrared observations). However, uncertainties in estimating the CO cloud masses are large since this requires knowledge of cloud velocities and the conversion factor between CO and molecular hydrogen (which also hinders our CO observations towards Kepler, see below) and large-scale observations in the submm are sorely needed to differentiate between the remnant and foreground.

Krause et al. (2004) re-reduced the SCUBA submm data and used Spitzer to observe Cas A at 160 μm (Hines et al. 2004), with OH absorption emission. They used the 'median' option in removing the baseline for the 850 μm data. Their final image has large negative features, which are of a greater level than the positive emission in the south. The SNR in this case is sitting on a negative background in the north and a positive background in the south. The measurement of the level of flux in the south is critical to their argument that this emission is from foreground material. They found a high degree of correlation with the dust emission and OH absorption seen towards the remnant (which incidentally is only detectable in absorption in the region strong in radio emission i.e. the bounded shocks). The correlation is less convincing when compared to the CO emission since this is has much larger, diffuse structure. They conclude that all of the submm emission is from foreground clouds with total dust mass. Their work suggests that no more than 0.2 $M_{solar}$ of dust can be associated with the ejecta itself, although this is almost a factor of 3 higher if the 'normal' dust absorption coefficient value for the diffuse ISM is used when estimating the dust mass in the molecular cloud i.e ~ 0.07 $m^2$ $kg^{-1}$ (Krause et al. (2004) use a value of 0.18 $m^2$ $kg^{-1}$). Thus the amount of dust from the supernova Cas A, or its progenitor star is very uncertain. The correlation between submm continuum and molecular emission suggests that some of the dust may be indeed be foreground material that has been swept up by the blast wave, but the lack of a perfect correlation between the two suggests that some of the dust is made in the supernova/massive star wind. We note that a substantial correction for foreground material would bring the amount of dust in Cas A better in line with the mass of dust estimated for Kepler's SNR.

Given the controversy about the amount of dust in the Cas A SNR, it is important to determine whether or not Kepler's remnant is also contaminated by foreground material. Kepler's SNR was observed in the CO(J=2-1) line with the A3 receiver on the JCMT in 2004, as part of the JCMT Service Programme. We used the wide-band mode of the DAS spectrometer, which has a bandwidth of 1.8 GHz and spectral resolution of 1.97 km/s. We mapped a square region 6 x 6 arcmin. Details about the reduction process will be given in Gomez et al. (in preparation). We made no significant detection of CO over the entire range -150 < v < 150 km/s. Given that there is no detected signal, we can only calculate a 3σ upper limit from the maps over the relevant velocity range of possible CO clouds.

Kepler's SNR is far from the Galactic plane, suggesting that confusion from foreground clouds should be easily seen and features associated with the remnant should also be easily recognisable (see for example a recent HI study towards the remnant, Reynoso & Goss (1999)). At lower Galactic latitudes, there is evidence for molecular cloud structures with a wide velocity range between -20 < v < +40 km/s although 90 % of the CO emission in this region is within the velocity range -10 < v < +20 km/s (Dame et al. 2001). However, a CO latitude-velocity map at a Kepler's location, shows that the velocity range of clouds at the higher latitude of Kepler's remnant is roughly -5 < v < +10 km/s (kindly provided by T. Dame, private communication) with most of the emission confined to the smaller range 0 < v < +5 km/s (see also Fig 5(a) in Dame et al. 2001). Figure 9 shows the CO emission towards Kepler's SNR over the velocity range -5 < v < +5 km/s with the 850 μm contours and the location of the shock front overlaid. Dust clumps are labelled A – E. We can determine the gas mass in the CO data using our upper limit I(CO),

$$M_{H_2} = 4.49 \pi R^2 I(CO)$$
$$M_{gas} = 1.36 M_{H_2}$$

Integrating the CO intensity over a wide velocity range of $\Delta v \sim 30$ km/s, we estimate that the 3σ upper limits on the dust-to-gas ratios in the clouds A-E is 48, 44, 20, 58, 34. This is lower than the nominal 100 – 200 values seen in the interstellar medium suggesting that the dust is not from a foreground molecular cloud. However, these values depend on the velocity width of clouds which could conservatively increase up to $\Delta v \sim 60$ km/s. In this case, the dust to gas ratios would increase by a factor of $\sqrt{60/30}$. Ironically, not detecting a clear signal from CO gas means we have to integrate over all possible clouds at this location to be conservative. The likelihood of a 60 km/s cloud is remote unless it is physically interacting with the remnant. The typical size of molecular clouds with $\Delta v \sim 60$ km/s. is greater than 30 pc (Solomon et al. 1987), ten times larger than SCUBA clumps. Given the errors involved in the velocity width in addition to the errors in estimating the dust mass of the clouds, we require deeper observations of CO emission towards Kepler to obtain far better sensitivity.

## Conclusion

The question of exactly how much dust is formed in supernovae is still controversial. Chemical evolution models suggest the need for a supernova (or rapidly-evolving) source of dust in both the early Universe and in our own Galaxy.  In 2003, the first observational evidence of copious amounts of dust in supernova remnants was provided by SCUBA, probing the emission from cold dust, dust that previous far-infrared telescopes had missed.  Other explanations for the submillimetre emission were put forward, namely (1) the emission seen in SCUBA was from 'exotic' iron needles which are efficient at radiating in the submillimetre and (2) that the emission was actually from interstellar clouds and not the supernovae themselves.  Recent work suggests that one of the remnants, Cassiopeia A, is contaminated by emission from dust in a foreground cloud, and although our observations of carbon monoxide emission from gas towards Cas A confirm that some of the dust emission may be from foreground material, this does not explain why the dust peaks in our SCUBA image fall between the bowshock and the reverse shock.  This of course could be a chance alignment, but the coincidence is evidence that some of the dust was either formed in the supernova or swept up from the surrounding ISM.  Our limited observations of carbon monoxide emission towards Kepler suggest so far, that the SCUBA emission is from dust in the remnant and not foreground.  There are many uncertainties when estimating gas masses from CO data, which need to be considered before this result can be verified/disproved.  We conclude that even if most of the dust in Cas A and Kepler is foreground, we may still be left with 0.1 $M_{solar}$ of dust formed by the pre-supernova massive star or in the supernova blast wave.  This is more than enough to explain dusty galaxies at high redshifts and solves the dust budget crisis in our own Galaxy.  We eagerly await results from SCUBA-2 and the Herschel Space Observatory which will finally have the combined resolution, mapping speeds and sensitivity to resolve the question of the origin of dust in the Universe.

# Figures

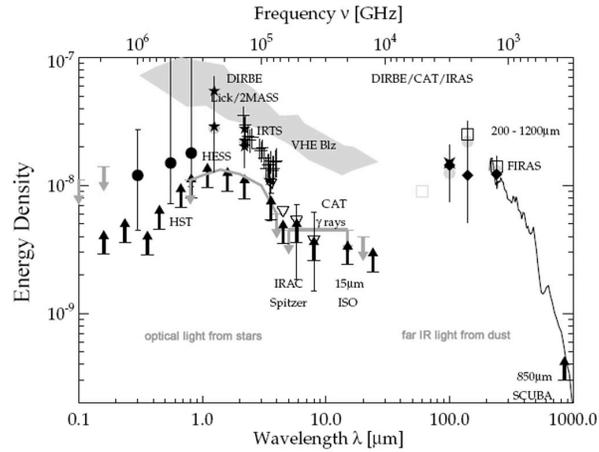

Fig 1: The integrated background energy in the optical-infrared regime (adapted from Dole et al. 2006). Starlight and reprocessed starlight contribute almost equal amounts to the background energy.

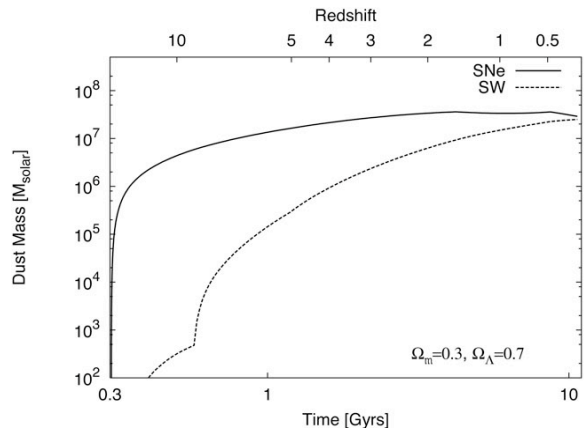

Fig 2: The dust evolution of a galaxy ($10^{10}$ $M_{solar}$) with time with star formation rate $1 M_{solar}$ yr$^{-1}$ (Morgan & Edmunds 2003). Redshifts are calculated using the concordance cosmological model, $\Omega_m = 0.3$ and $\Omega_\Lambda = 0.7$. In this case the contributions from dust formation in supernova (SNe) using theoretical estimates (e.g. Todini & Ferrara 2001) and observations of stellar winds (SW) (Whittet 2003) are shown separately as solid and dotted lines respectively.

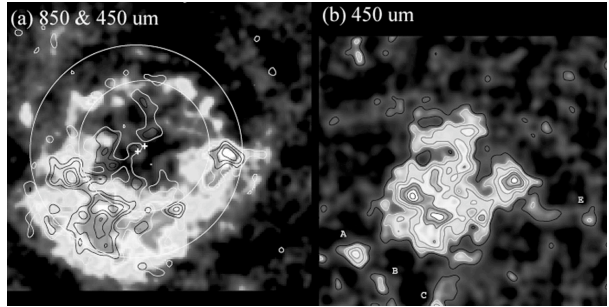

Fig 3. SCUBA images of the Cassiopeia A supernova remnant at (a) 850 μm and (b) 450 μm (from Dunne et al. 2003b). In (a) we show the synchrotron-subtracted 850 μm image with 450 μm contours overlaid (3σ + 1σ). The forward and reverse shock fronts seen in the X-ray from the supernova blast wave are overlaid (Gotthelf et al. 2001).

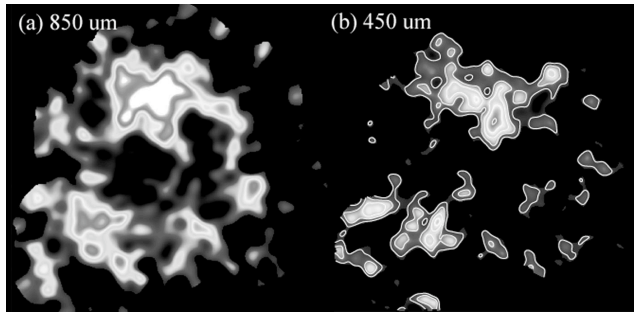

Fig 4. SCUBA images of Kepler's supernova remnant at (a) 850 μm and (b) 450 μm (Morgan et al. 2003). The original reduced data maps have been divided by a simulated noise map, these images therefore represent a signal-to-noise map of Kepler.

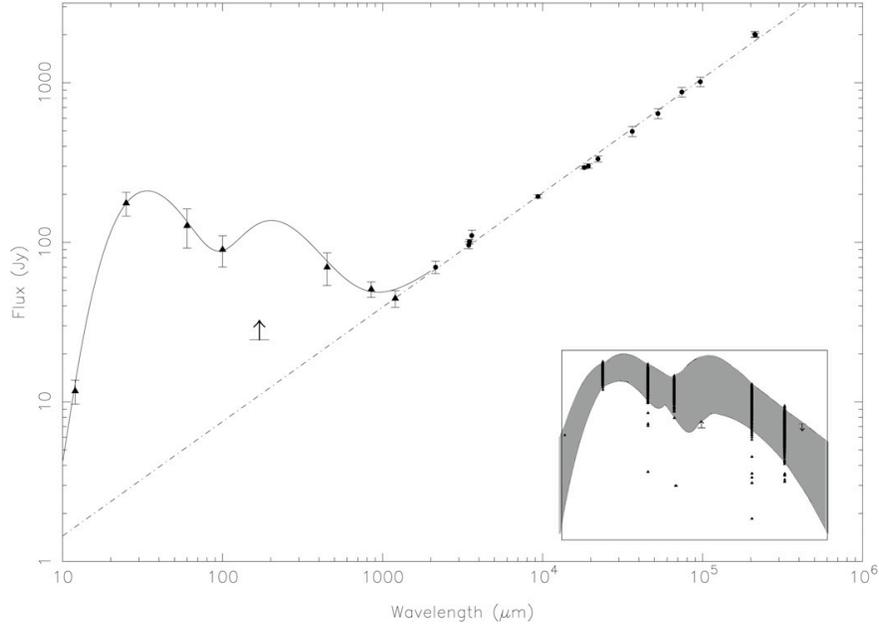

Fig 5. The spectral energy distribution of Cas A. The solid lines represent the two-temperature-best-fit $\chi^2$ test, with dot-dashed lines representing the hot and cold component of dust grains in each SED. The 170 μm lower limit from ISO is shown (Tuffs et al. 1999). Inset we show the 3000 fits from the bootstrap technique from the original data points (those with $\chi^2 < 3.0$ are shown). Although there is no data point to confirm the existence of the second peak around 200 - 400 μm, it appears to remain even when fitting extremes to the SED. Best fit parameters are $T_{hot} \sim 112\,^{+11}_{-21}$ K, $T_{cold} \sim 17 \pm 3.6$ K, $\beta \sim 0.9\,^{+0.8}_{-0.6}$ with radio power law of $\nu^{-0.61}$.

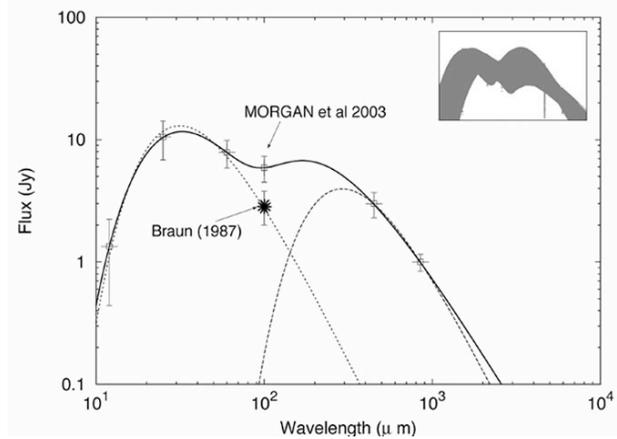

Fig 6. The spectral energy distribution of Kepler. Two 'best' fits to the IR-submm data for Kepler are shown due to uncertainty in IRAS fluxes for Kepler at 100 μm ranging from 2.9 Jy (value from pointed observations of the remnant) - 5.9 Jy (the average of all values for Kepler published with IRAS). Neither fit rule out a cold dust component. Inset are the results from the bootstrap technique. Best fit parameters are $T_{hot}$ ~ 102 ± 12 K, $T_{cold}$ ~ $17^{+2}_{-3}$ K, β ~ 1.2 ± 0.4 with radio power law of $v^{-0.71}$.

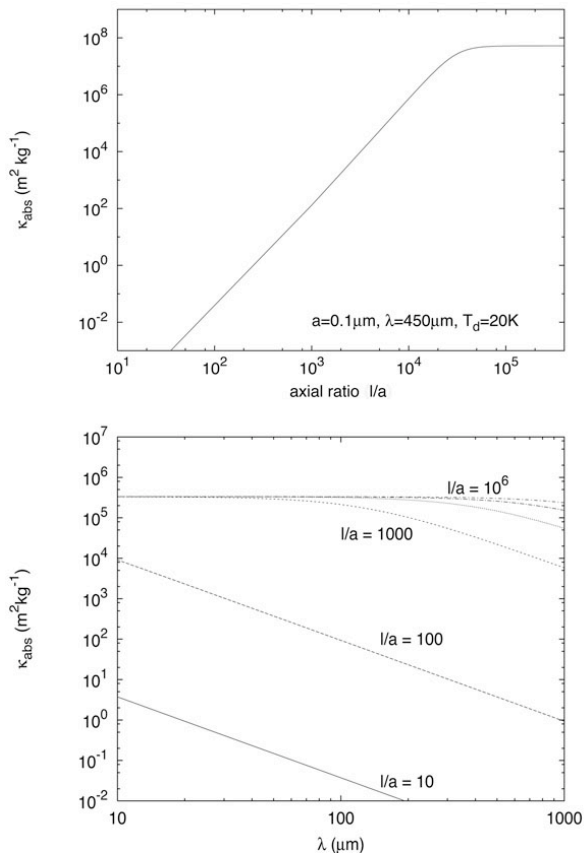

Fig 7. (a) The variation of the absorption coefficient of an iron needle with axial ratios ($l/a$) for typical interstellar grain size, a = 0.1 μm, at 450 μm (following Li 2003). (b) The variation of the absorption coefficient of iron needles with wavelength for different axial ratios ($l/a$). The needles are modelled as antenna with resistivity $10^{-5}$ Ω cm. (Gomez et al. 2005).

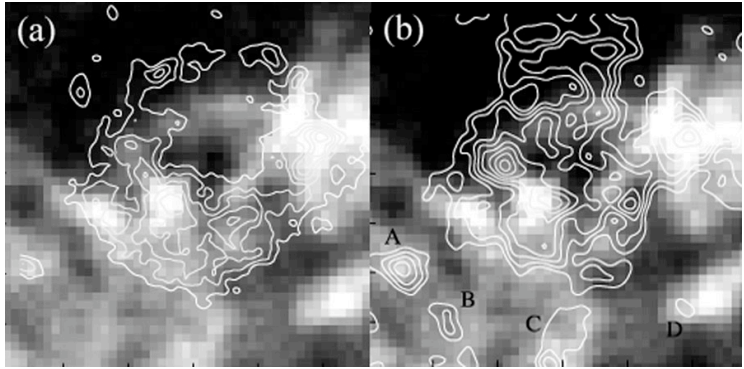

Fig 8. Integrated CO emission towards Cassiopeia A over the velocity interval $-50 < v < -35$ km/s, which includes all the gas from the Perseus spiral arm. Overlaid are SCUBA contours at (a) 850 μm with synchrotron emission subtracted and (b) 450 μm.

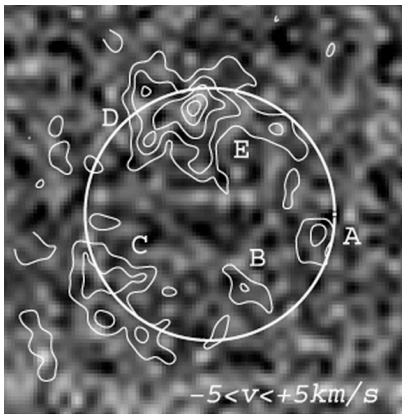

Fig 9. Integrated CO emission towards Kepler's supernova remnant over the velocity interval $-5 < v < +5$ km/s. The 850 μm SCUBA contours are shown ($3\sigma + 2\sigma$) along with the position of the forward shock as traced by X-ray observations. No signal is detected in these images with $3\sigma$ upper limit I(CO) ~ 2.21 K km/s. Cold dust clumps are labelled A – E.